\documentclass[reprint,amsmath,amssymb,aps,prstper,nofootinbib,superscriptaddress, longbibliography]{revtex4-2}

\usepackage{graphicx}
\usepackage{dcolumn}
\usepackage{bm}
\usepackage{comment}
\usepackage{cleveref}
\usepackage{subcaption}
\usepackage{wrapfig}
\usepackage{dblfloatfix}

\begin{document}

\preprint{APS/123-QED}

\title{Click, Watch, Learn: The Impact of Student Self-Study Materials on Physics E\&M Course Outcomes}

\author{James K. Hirons}
 \affiliation{Department of Physics \& Astronomy, Texas A\&M University, College Station, Texas, 77843}
\author{Jonathan D. Perry}
\affiliation{Department of Physics, University of Texas at Austin, Austin, Texas 78712}
\author{Dawson T. Nodurft}
\affiliation{Department of Physics \& Astronomy, Texas A\&M University, College Station, Texas, 77843}
\author{Scott Crawford}
\affiliation{Department of Statistics, Texas A\&M University, College Station, Texas, 77843}
\author{William Bassichis}
 \affiliation{Department of Physics \& Astronomy, Texas A\&M University, College Station, Texas, 77843}
 \author{Tatiana L. Erukhimova}
 \email{etanya@tamu.edu}
 \affiliation{Department of Physics \& Astronomy, Texas A\&M University, College Station, Texas, 77843}

\date{\today}

\begin{abstract}
Performance in introductory courses, particularly physics, is often crucial for student success in STEM majors and can impact an individual’s tendency to persist in their chosen field. To enhance students' individual learning experiences, faculty at many universities have worked to develop open-access, self-study materials to help build conceptual understanding and problem-solving skills. Faculty at Texas A\&M University have contributed to these efforts, creating more than 200 online video resources and a broad bank of prior exams available to students. This work explores and measures the impact that these resources can have on student course outcomes in an introductory, calculus-based electricity and magnetism course. Data were collected from three fall semesters, 2021-2023, including classroom performance, a conceptual assessment, and relevant university level data to contextualize student background and pre-class abilities. We present results from a regression analysis examining the relative importance of multiple resources offered concurrently to students. We also discuss how the use of supplemental materials affects course performance, alongside other potential predictors such as demographics and prior preparation. Similar to prior studies, we found that relevant prior preparation in mathematics was the strongest predictor of student performance, with prior physics knowledge being a weaker but still statistically significant predictor. Students' utilization of supplemental old exams was the second highest predictor of student performance. First generation students were observed to have a slightly lower average performance on exams. However, interaction terms in the regression models indicated that first generation students using supplemental old exams were able to close this gap. Through anonymous surveys, students reported warm impressions of the materials, with over 80\% of students sharing that they had a noticeable contribution to their learning outside of the classroom and 98\% stating they would recommend them to their peers.

\end{abstract}

\maketitle

\section{\label{sec:level1}Introduction}
Introductory physics courses play a critical role in building foundational knowledge for students pursuing science, technology, engineering, and mathematics (STEM) majors. Calculus-based physics, in particular, is a required component for all physics and engineering students during their first two years of college. As part of students’ early academic experiences, these courses are often pivotal to their success and progression within their chosen fields \cite{ReportCard2011, good2012women}. A recent five-year longitudinal study on attrition and persistence in undergraduate physics programs found that the majority of students who lose interest in completing a physics major make that decision during their first or second year \cite{AIP2024retention}, underscoring the importance of student experiences in introductory physics courses. Research shows that success in introductory physics courses is strongly linked to students’ self-efficacy in math and physics, prior academic preparation, and access to effective learning resources \cite{ lent1984persistence, lent1987predicting, Lindstrom2011, sadaghiani2011, ballen2017, salehi2019, Stewart2021, Iwuanyanwu2022, meltzer2025pre}. While strong mathematical foundations are consistently associated with better performance, incoming students often display a wide range of preparedness levels \cite{Sadler2001HSprep, kost2009characterizing, meltzer2025pre}.

Compounding these challenges, introductory courses are frequently taught in large-enrollment settings, where classes have heterogeneous levels of student preparation, ranging from those with substantial preparation or high scores on Advanced Placement exams to those who may have never had a prior physics course. Due to the size of these courses, there are often limited opportunities for individualized attention and tailored support. Prior literature shows that first-generation students can experience additional academic challenges, compounding the issues of large classes, related to academic self-regulation \cite{stebleton2012, falcon2015}. In recent decades, a substantial body of knowledge has evolved offering instructors myriad strategies for effective teaching \cite{schwartz1998time, NRC2012understanding, NRC2013adapting, mestre2020science}. While these strategies have made significant strides in improving student learning outcomes across many introductory courses, there remains room for improvement beyond changes to classroom pedagogy. It is important to recall that a lot of student learning still takes place outside of the scheduled class time, with students spending sometimes considerable effort preparing for lectures, completing assignments, and studying for exams---often without direct access to guidance from the instructor or instructional team \cite{mestre2020science}. In today's digital landscape, students have become increasingly reliant on online resources to support their learning \cite{dolch2021higher}.

Online platforms such as FlipItPhysics and Interactive Video Vignettes provide exemplars of efforts that have demonstrated measurable benefits for student learning and performance by offering structured, interactive content that reinforces core STEM concepts \cite{stelzer2009, Laws2015vignettes, moore2018efficacyMLM}. FlipItPhysics, for example, uses pre-lecture videos to replace traditional lectures, allowing classroom time to focus on problem-solving and demonstrations. Broadly supplemental learning materials---including structured videos, problem-solving guides, and archived exams---have emerged as valuable tools for bridging gaps in student preparation \cite{chen2010multimedia, moore2018efficacyMLM, sadaghiani2011, Fakcharoenphol2011, hill2015, perry2019newresource}. Prior research has shown that supplemental resources can aid in students' development of conceptual understandings in physics \cite{chen2010multimedia, sadaghiani2011, hill2015}. In addition, interactive problem solving tools have been shown to improve participation and improve students' ability to navigate complex problems \cite{Singh2010olm, Choirudin2025}. However, the impact of these materials can vary based on individual interaction patterns and study strategies. 

The increased availability of digital course materials during the pandemic proved valuable to students. When surveyed, many expressed a desire to retain access to recorded lectures and online content even in post-pandemic, face-to-face classrooms \cite{Wilcox2020experience, Dew2021Changes}. To support student success, address gaps in preparation, and accommodate diverse learning needs, instructors at many institutions have developed open-access resources that foster self-directed learning outside the classroom \cite{stelzer2009, Singh2010olm,  hill2015, Hill2017,perry2019newresource}. As the field of physics education evolves, multimedia resources have become an increasingly prominent instructional tool, reshaping the ways students engage with complex material \cite{chen2010multimedia, cahill2014}. This is especially critical, as success in introductory science and mathematics courses can determine whether students persist through mid-level and advanced coursework \cite{ReportCard2011}.

In this study, we focused on the role of learning resources by examining how multiple supplemental learning resources influenced student success in a large, calculus-based introductory electricity and magnetism (E\&M) course at Texas A\&M University.  Starting more than 10 years ago, physics faculty there have created a series of online supplemental materials to support student learning. We hypothesized that usage of our department’s supplemental materials can have a significant, measurable improvement on student course outcomes. Our specific research aims were to investigate three key questions: (1) what impact did supplemental resources have on student outcomes in introductory E\&M course, (2) how did the impacts differ between first-generation and continuing-generation students, and (3) what were students’ perceptions of the impact of these resources. This work expands on previous studies which have investigated the effect of individual resources, such as online videos, on learning by broadly investigating how multiple resources, each designed to meet different self-study needs, affect students' course outcomes. In the following sections, we describe the course context, supplemental resources, and analysis methods, present results from several regression models and surveys, and discuss the implications of our results.

\section{Research Methods}
 
\subsection{Course description}

The data for this study were collected from students who took a second semester calculus-based E\&M class during the fall semesters of 2021–2023 at Texas A\&M University. These are large-enrollment courses which typically have around seven sections of \(\sim\)140 each, with an additional honors section around 50 students each semester. These courses are taught by a large team of instructors, usually having five or six faculty and upward of twelve graduate teaching assistants (TAs). All students used the same textbook, \textit{Don’t Panic: A Guide to Introductory Physics For Students of Science and Engineering}, Volume 2, by William Bassichis, a co-author of this paper. In addition to three hours of lecture per week, students attended an eighty-minute recitation session each week of no more than twenty-eight students. The recitation portion, where students practiced problem-solving both individually and in groups, was taught by TAs. 

All instructors (faculty and graduate students) met weekly to gather feedback from the previous week and discuss plans for the following week, ensuring instructional uniformity for students taking the course. Instructors utilized in-class lecture quizzes, either on paper or through the i-Clicker system, as well as weekly quizzes administered through Gradescope. At the beginning and end of the semester, students took the Brief Electricity and Magnetism Assessment (BEMA) \cite{BEMA}, completion of which was a small bonus towards their overall quiz grade, for the course equivalent to one weekly quiz. The course summative assessments included three common midterm exams and a comprehensive final exam. All exam questions were free-response, multi-step problems, with students expected to provide detailed solutions that indicated their thought process: the law(s) they used to approach the problem and the steps that led to the final answer.

The three midterm exams were common exams administered to all students taking the class at the same time. All instructors and TAs graded student papers immediately following the exam and returned them during the next lecture, ensuring that feedback was delivered while the content remained fresh in students’ minds---a practice shown by prior research to support students' learning \cite{mestre2020science}. To ensure uniformity, each problem on the exam was graded by the same team, consisting of a lead faculty member and TAs. The final comprehensive exams were written by individual instructors and these exams were scheduled on different days. However, all instructors submitted their final exams to a course coordinator to check for uniformity between instructors.

Students were encouraged to use the open-access supplemental materials \cite{tamuIntroPhysics} created over the years to assist learners with varying backgrounds and levels of preparation in mastering the course learning outcomes outside the classroom. These self-study materials were posted on the departmental website and were created with support from internal grants. Students were not given any course credit for their usage of supplemental materials. The open-access materials included four distinct resources tailored to different types of learners.

The resources consisted of the following:
\begin{enumerate}
        \item Chapter Outline Videos, developed by authors WB and TE, were designed to offer a structured overview and summary of key topics in each chapter of the required textbook, Don't Panic. These videos were intended for students to watch before reading the chapter and attending lectures.
        \item Conceptual and Example Videos, developed by WB, TE, and JP, were created to reinforce key topics after they had been introduced in class and to emphasize their application to common and important problems. These short videos included explanations, derivations, and demonstrations to help illustrate the concepts.
        \item Problem-solving Videos, also referred to as Recitation Videos, developed by TE, provided detailed solutions to multi-step problems that students might encounter in their homework or major summative assessments. Their in-depth approach was designed to help students develop and refine their problem-solving skills and techniques.
        \item Collection of midterm and final exams from previous years,  developed by WB and TE, spanning over fifteen years. The midterm exams provided only the answers, without solutions to the problems, while the final exams included detailed solutions.
\end{enumerate} 

\subsection{Data description}

To investigate the impact of supplemental materials on student learning, we collected course-level, departmental-level, and university-level data for this study. The course-level data included students' grades, out of a possible 100, on midterm exams, performance on final exams, their final grades, and BEMA performance measured during the first and last days of recitation. Prior preparation in physics was measured using the BEMA pre-score, while exam performance and final grades granted numeric and categorical representations of student success. Anonymous surveys were also conducted after each of the three midterm exams aimed at gauging self-reported student impressions of the impact and utility of the supplemental materials on their course performance.

The departmental-level data included student usage of the supplemental resources, which was tracked by monitoring student clicks on each individual embedded hyperlink on our supplemental material website. These counts provided an estimate of each student's usage of individual videos or past exams and enabled us to create four additional explanatory variables: one for each type of video and one for the past exams. 

The university-level data included demographics and background preparation in mathematics. The demographic data included information related to student gender, ethnicity, and first-generation student status. Prior preparation in mathematics was assessed using the quantitative portion of the Scholastic Aptitude Test (SAT-math), the student's final letter grade in the first course of the calculus sequence taken at Texas A\&M, and their performance on the required university-administered Math Placement Exam (MPE). Letter grades from the students' calculus courses were treated on a standard 4 point scale, with A\(\equiv\)4, B\(\equiv\)3, etc. For uniformity in the analysis, these numeric grade values, along with the other three prior preparation variables, were normalized as z-scores \cite{StatPower}. All three data sets were matched for each student by the university assessment team, anonymized, and provided to the researchers. 

\subsection{Statistical Methods}
\begin{figure}[h!]
    \begin{subfigure}{0.5\textwidth}
        \includegraphics[width=\linewidth]{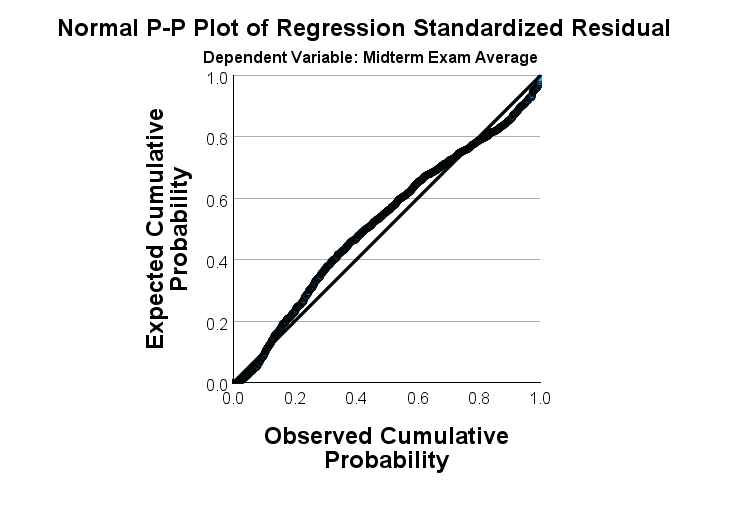}
        \caption{}
        \label{fig:p-p plot}
    \end{subfigure}     
    \begin{subfigure}{0.5\textwidth}
        \includegraphics[width=\linewidth]{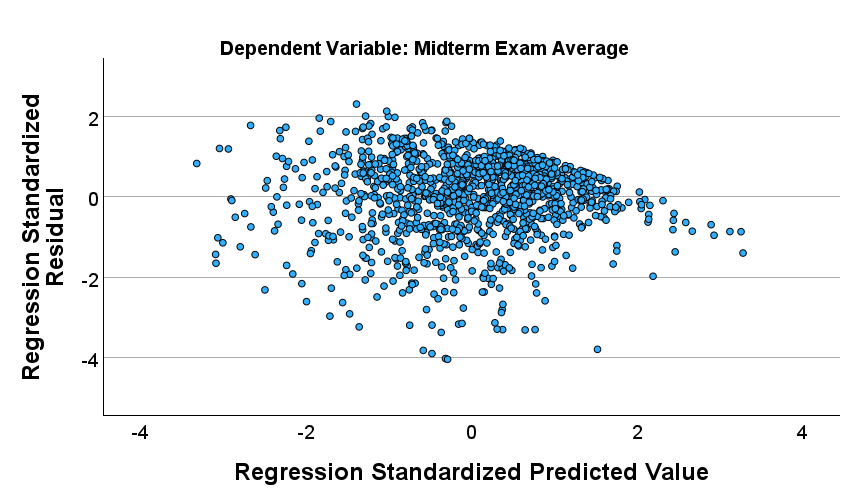}
        \caption{}
        \label{fig:residuals}
    \end{subfigure}
    \caption{Plots for check of assumptions for regression models, including (a) p-p plot of residuals, and (b) a scatterplot of residuals. }
\end{figure}

Multiple linear regression analysis was conducted using the statistical software package SPSS. Prior to interpreting the results, key assumptions underlying the regression model were systematically evaluated: normality of residuals, homoscedasticity, and absence of multicollinearity. To assess the normality assumption, a P-P plot comparing the cumulative distribution functions of the observed and expected residuals was generated. The results pertaining to normality and homoscedasticity are presented in Fig. 1. The results fall within acceptable threshold for normality, Figure \ref{fig:p-p plot}. Homoscedasticity was examined via a scatter plot of standardized predicted values against standardized residuals, Figure \ref{fig:residuals}. This assumption was met, though minor clustering was observed, attributable to the ceiling effect of exam scores, which limit average values to a maximum of 100. Assumptions were also checked for each independent, non-binary variable included in regression models. These plots, shown in the Supplemental Materials \cite{supp}, indicate that assumptions were met for each parameter. To evaluate multicollinearity, Variance Inflation Factors (VIF) were computed for all explanatory variables across the regression models. All VIFs were $\leq$ 1.7, which is well below the conventional threshold of 4 \cite{PennStateStats}, indicating that multicollinearity did not impact estimated regression coefficients.

\begin{figure}[h!]
    \begin{subfigure}{0.5\textwidth}
        \includegraphics[width=\linewidth]{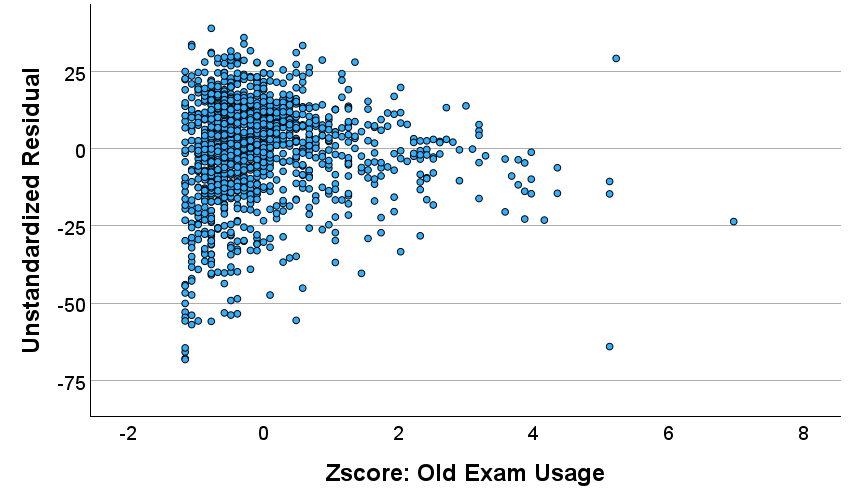}
        \caption{}
        \label{fig:OldExamNoMod}
    \end{subfigure}     
    \begin{subfigure}{0.5\textwidth}
        \includegraphics[width=\linewidth]{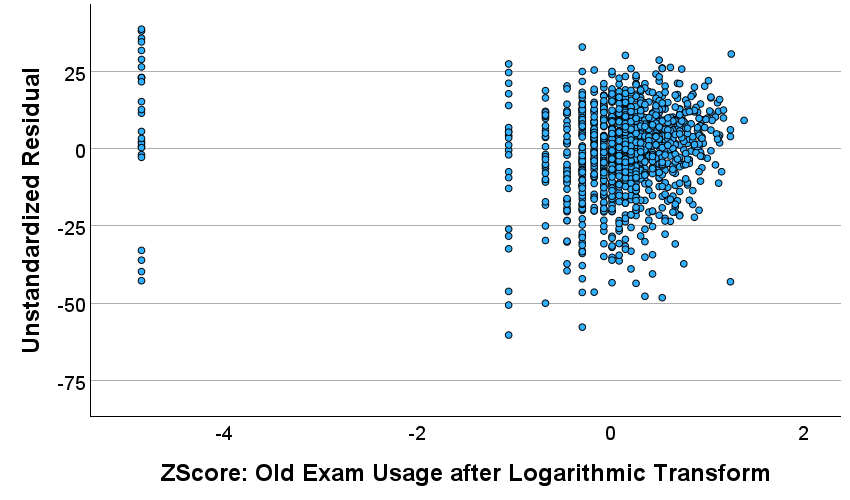}
        \caption{}
        \label{fig:OldExamLog}
    \end{subfigure}
    \caption{Scatterplots of residuals for usage of supplemental old exams (a) No modifications, and (b) Post logarithmic transform. }
\end{figure}

Variables expressed as counts are often skewed due to the inherent lower bound of zero; for instance, a student cannot engage with a video fewer than zero times. This limitation is reflected in the residual patterns observed for the supplemental resource variables, Fig. \ref{fig:OldExamNoMod}. Consequently, the explanatory variables corresponding to supplemental resources demonstrated non-normality when analyzed in linear space. To mitigate this issue, a logarithmic transformation was applied to these variables. Post-transformation residual plots revealed that the assumption of normality was largely satisfied, aside from observable clustering at the lower bound, Fig. \ref{fig:OldExamLog}.

In interpreting results of multiple regression models, several key metrics help in evaluating both the overall model performance and the relative influence of individual predictors. R-square values \( 0 \leq R^2 \leq 1 \) quantify the proportion of total variance explained by the model. Beta coefficients represent the relative contribution of each independent variable to the observed effect. A threshold of statistical significance of $p<$0.05 was adopted, a regularly selected value for such studies \cite{StatPower}.

Initial regression models incorporated all available explanatory variables. To refine the model, we then employed backward elimination \cite{LinearModelsWithR}, which reduced the number of explanatory variables until only statistically significant factors remain. These explanatory factors included demographics (student gender, ethnicity, and first-generation status), usage metrics for each of the four supplemental materials, and indicators of prior academic preparation: SAT quantitative score, BEMA pre-test score, score on the university-administered MPE, and the final letter grade earned in Calculus I. 

A total of 2556 students were enrolled over the three semesters, 2124 of which opted into this study. However, due to missing data (e.g. BEMA pre, calculus grade, first generation status) we had complete data for 1368 students. As final exams and final course grades were assigned by individual instructors, the most uniform measure of student performance within the course was the common midterm exams. Therefore, the response variable in our multiple regression models was the average score between all three midterm exams. We did not include the BEMA post-test as a response variable due to low participation rates. 

To examine impacts between multiple explanatory variables, we employed a limited number of interaction terms. These interaction terms facilitated predictions for specific student subgroups and enabled testing of the hypothesis that outcomes may differ by, for instance, generational status \cite{kost2009characterizing, lunt2015interactions, Stewart2021}. Interaction terms were created by multiplying the value of each relevant variable by a binary indicator of first-generation status. For example, the interaction term for student engagement with supplemental old exams captured usage counts exclusively for first-generation students, with values set to zero for continuing-generation students.

\section{Results}
Preliminary regression models incorporated student ethnicity as an explanatory variable, revealing a statistically significant negative correlation between exam performance and students identifying as Hispanic or Asian. No significant effects were identified for the remaining ethnic groups. This pattern aligned with previous research indicating that observed performance disparities between minority and majority students may be attributable to other factors, particularly pre-course academic preparation \cite{brewe2010, Stewart2021}. Subsequent analysis suggested that the statistical significance associated with ethnicity was predominantly driven by first-generation student status within these broader categories. In response to this finding we omitted ethnicity from further analysis. 

This section discusses three models. The first model was a linear regression that included interaction terms linking first-generation student status to each of the four supplemental resources, with all variables represented in linear space. The second model applied logarithmic transformations to variables exhibiting non-normality, as previously noted. The final model combined interaction terms and logarithmic transformations to enable a more in-depth analysis of factors contributing to student performance.

\subsection{First Model: Interaction Terms}

To explore the interaction between students' first-generation status, their use of supplemental materials, and academic performance, we employed multiple linear regression. Interaction terms met all assumptions for regression modeling. Descriptive statistics and results for this model are presented in Table \ref{tab1}. Our sample consisted of 1368 students, 12.2\% of whom had first-generation status. 

\begin{table}[h!]
    \caption{\label{tab1} Descriptive statistics and results for multiple regression model including interaction term between first-generation status and usage of supplemental materials.}
    \centering
    \begin{tabular}{l   r   r}
    \hline
    \hline
    \multicolumn{3}{c}{Descriptive Statistics for Interaction-Term Model, N = 1368}\\
        \hline
        Variable & Mean & STD\\
        \hline
        Exam Average & 75.676 & 17.472\\
        Math Placement & 0.073 & 0.868\\
        Calculus 1 Grade & 0.072 & 0.953\\
        BEMA Pre & 0.000 & 0.979\\
        Outline Videos & 0.005 & 1.020\\
        Old Exams & 0.046 & 1.008\\
        First Gen (Binary) & 0.122 & n/a\\
        First Gen x Old Exams & 0.020 & 0.445\\
        \hline \\
        \hline
        \hline
        \multicolumn{3}{c}{Multiple Regression Model including Interaction Term} \\
    \multicolumn{3}{c}{ R Square = 0.302}\\
    \hline
    Independent Variable & Beta & Significance\\
        \hline
        Calculus 1 Grade & 0.327 & $<$0.001\\
        Old Exams & 0.221 & $<$0.001\\
        Math Placement  & 0.175 & $<$0.001\\
        BEMA Pre & 0.157 & $<$0.001\\
        FirstGen x Old Exams & 0.050 & 0.047\\
        First Gen (Binary) & -0.066 & 0.004\\
        Outline Videos & -0.084 & $<$0.001\\
        \hline
    \end{tabular}
\end{table}

The model has an R-square of 0.302, indicating that it took into account 30.2\% of the variability in student performance. The strongest predictor of student performance was their Calculus 1 grade $(\beta$ = 0.327). The second strongest predictor was the use of supplemental old exams ($\beta$ = 0.221). Other significant prior preparation variables included MPE scores ($\beta$ = 0.175) and BEMA pre-test scores ($\beta$ = 0.157). A small negative correlation was observed between first-generation status and course performance ($\beta$ = -0.066). However, a positive correlation was found for the interaction term between first-generation status and usage of supplemental old exams ($\beta$ = 0.050). Among the supplemental videos, only the Chapter outline videos showed a statistically significant effect, with a negative correlation ($\beta$ = -0.084).

\subsection{Second Model: Logarithmic Transformations}

Due to the observed non-linearity within the residual plots for all supplemental materials, we then implemented a logarithmic transformation to better understand the relationship between resource utilization and course performance. As shown in Table \ref{tab3}, the sample size for this model is $N$ = 1402. The slight increase in the $N$ is due to the removal of first-generation status as this identifier was missing from 34 students in the data base. The first-generation status variable showed no statistical significance within this model and was therefore removed from the reduced model during backward elimination. 

\begin{table}[h!]
    \caption{\label{tab3} Descriptive statistics and results for multiple regression model including logarithmic transformation of supplemental materials explanatory variables.}
    \centering
    \begin{tabular}{l   r   r}
    \hline
    \hline
    \multicolumn{3}{c}{Descriptive Statistics for LOG Transform Model, N = 1402}\\
        \hline
        Variable & Mean & STD\\
        \hline
        Exam Average & 75.809 & 17.454\\
        Math Placement & 0.078 & 0.869\\
        Calculus 1 Grade & 0.080 & 0.953\\
        BEMA Pre & 0.0004 & 0.976\\
        Outline Videos* & -0.009 & 1.001\\
        Old Exams* & 0.092 & 0.825\\
        \hline
        \\
        \hline
    \hline
    \multicolumn{3}{c}{Multiple Regression Model including Logarithmic Transform} \\
    \multicolumn{3}{c}{R Square = 0.314}\\
    \hline
    Independent Variable & Beta & Significance\\
        \hline
        Calculus 1 Grade & 0.325 & $<$0.001\\
        Old Exams* & 0.262 & $<$0.001\\
        Math Placement  & 0.176 & $<$0.001\\
        BEMA Pre & 0.145 & $<$0.001\\
        Outline Videos* & -0.106 & $<$0.001\\
        \hline
    \end{tabular}
\end{table}

Within this multiple regression model, seen in Table \ref{tab3}, we had an R-Square of 0.314, which implies that we accounted for 31.4\% of the variability. This model showed that a student's previous success in calculus was the highest predictor of student performance in introductory E\&M ($\beta$ = 0.325). Also, as previously seen, student usage of supplemental old exams was the second highest predictor of student performance ($\beta$ = 0.262). Two other measures of prior preparation, MPE ($\beta$ = 0.176) and BEMA ($\beta$ = 0.145), also showed positive correlation with students' performance on exams. Student usage of chapter outline videos showed a negative correlation ($\beta$ = -0.106) with student performance, while the supplemental review and recitation-style problem-solving videos showed no significance, similar to what was observed in the first model. 

\subsection{Third Model: Hybrid Methods}
In the first model, we found that both first-generation status and the use of supplemental materials by first-generation students were statistically significant predictors of students' average midterm scores. However, in the second model first-generation status no longer showed statistical significance. To obtain a more complete picture of the key predictors of student success in this course, we built a hybrid-model that combined both approaches, specifically interaction terms and logarithmic transformations for the supplemental materials variables. Descriptive statistics and results are shown in Table \ref{tab5}. For this model we had a sample size of 1368, 12.2\% of which were first-generation students. 

\begin{table}[h!]
    \caption{\label{tab5} Descriptive statistics and results for multiple regression model including logarithmic transformation of supplemental materials explanatory variables and first-generation interaction term.}
    \centering
    \begin{tabular}{l   r   r}
    \hline
    \hline
    \multicolumn{3}{c}{Descriptive Statistics for Hybrid Model, N = 1368}\\
        \hline
        Variable & Mean & STD\\
        \hline
        Exam Average & 75.676 & 17.472\\
        Math Placement & 0.073 & 0.868\\
        Calculus 1 Grade & 0.072 & 0.953\\
        BEMA Pre & 0.0004 & 0.979\\
        Outline Videos* & -0.012 & 1.0003\\
        Old Exams* & 0.087 & 0.832\\
        First Gen (Binary) & 0.122 & n/a\\
        First Gen x Old Exams* & 0.232 & 0.923\\
        \hline
        \\
          \hline
    \hline
    \multicolumn{3}{c}{Hybrid Multiple Regression Model} \\
    \multicolumn{3}{c}{R Square = 0.313}\\
    \hline
    Independent Variable & Beta & Significance\\
        \hline
        Calculus 1 Grade & 0.314 & $<$0.001\\
        Old Exams* & 0.236 & $<$0.001\\
        Math Placement  & 0.173 & $<$0.001\\
        BEMA Pre & 0.147 & $<$0.001\\
        FirstGen x Old Exams* & 0.075 & 0.028\\
        First Gen (Binary) & -0.092 & 0.005\\
        Outline Videos* & -0.109 & $<$0.001\\
        \hline
    \end{tabular}
\end{table}

With an R-Square = 0.313, the predictive power of this hybrid model is comparable to prior models. Notably, a student's calculus grade remains the strongest predictor of exam performance ($\beta$ = 0.314), whereas usage of the supplemental old exams followed in second ($\beta$ = 0.236). Other variables representing prior preparation, MPE ($\beta$ = 0.173) and BEMA ($\beta$ = 0.147), were positively correlated with performance in our course. First-generation status was negatively correlated with average exam performance ($\beta = -0.092$), indicating that these students scored lower on average compared to their continuing-generation peers. At the same time, first-generation students' usage of prior years' exams showed a positive correlation ($\beta$ = 0.075) indicating that the observed differences could be counteracted through utilization of this resource. Student usage of chapter outline videos showed a negative correlation ($\beta$ = -0.109) with student performance, while the supplemental review and recitation-style problem-solving videos showed no significance. This echoes the results from our prior models.

\subsection{Student Perceptions Through Surveys}
Anonymous surveys were conducted to gauge student impressions about the utility and impact of our supplemental self-study resources. Data from all three semesters were compiled and student responses are discussed below. When asked about their physics experience from high school, Table \ref{tab7}, less than a quarter of students felt prepared for university-level physics. Meanwhile, roughly 2/3 of students said they did not feel prepared for this course, looking back on their high school experience. 

\begin{table}[h!]
    \caption{\label{tab7} Survey Question One: Did you take any physics classes in high school? If so, do you feel like they prepared you for this class? Compiled over all three semesters.}
    \centering
    \begin{tabular}{p{5cm}   r   r}
    \hline
    \hline
        Answer Choice & Exam 1 & Exam 3\\
        \hline
        Yes, and I do think they prepared me & 23.93\% & 23.76\%\\ 
        Yes, and I do NOT think they prepared me & 68.49\% & 66.14\%\\
        No, I did not take any high school physics classes & 7.58\% & 10.10\%\\
        \hline
    \end{tabular}
\end{table}

The next two questions focused on students' use of online resources to support their learning outside of class time, Tables \ref{tab8} \& \ref{tab9}. Almost all students, 99\%, indicated that at least a little of their outside learning came from online resources in general. Over 80\% of students reported that most of, if not all, of their outside learning of the material could be credited to our department's supplemental resources. From both of these survey questions, it can be seen that very few students rely purely on lectures and the textbook for their learning. 

\begin{table}[h!]
    \caption{\label{tab8} Survey Question Two: How much of your learning for this course, outside of class time, do online supplemental resources contribute to?}
    \centering
    \begin{tabular}{l   r   r   r}
    \hline
    \hline
        Answer Choice & Exam 1 & Exam 2 & Exam 3\\
        \hline
        All of it & 19.03\% & 27.65\% & 31.28\%\\        
        Most of it & 43.32\% & 45.22\% & 46.16\%\\
        Some of it & 28.69\% & 21.18\% & 18.69\%\\
        Only a little & 7.10\% & 4.81\% & 3.28\%\\
        None & 0.58\% & 1.14\% & 0.59\%\\
        \hline
    \end{tabular}
\end{table}

\begin{table}[h!]
    \caption{\label{tab9} Survey Question Three: Did the supplemental resources provided by the department noticeably contribute to your learning of the material outside of class time?}
    \centering
    \begin{tabular}{l   r   r   r}
    \hline
    \hline
        Answer Choice & Exam 1 & Exam 2 & Exam 3\\
        \hline
        All of the time & 43.09\% & 44.78\% & 51.65\%\\
        Most of the time & 37.52\% & 37.39\% & 35.11\%\\
        Some of the time & 14.40\% & 13.50\% & 10.67\%\\
        Rarely & 2.26\% & 2.71\% & 1.65\%\\
        Did not use & 2.73\% & 1.62\% & 0.92\%\\
        \hline
    \end{tabular}
\end{table}

Students were also asked which of the four types of supplemental resources they utilized, Table \ref{tab10}. Roughly half of students used the recitation-style problem solving videos, and over 30\% and 20\% of them used the supplemental concept videos and chapter outline videos respectively. A much higher percentage of students, over 82\%, made use of the repository of previous years' exams over the semester. Roughly half of students also reported that they made intentional changes to their study habits after using these resources, Table \ref{tab11}.

\begin{table}[h!]
    \caption{\label{tab10} Survey Question Four: Which of the following supplemental resources provided by the department did you use? (Select all that apply)}
    \centering
    \begin{tabular}{l   r   r   r}
    \hline
    \hline
        Answer Choice & Exam 1 & Exam 2 & Exam 3\\
        \hline
        Supplemental Videos & 32.63\% & 37.62\% & 34.79\%\\        
        Recitation Videos & 49.59\% & 54.16\% & 55.24\%\\
        Chapter Outline Videos & 26.88\% & 24.32\% & 22.83\%\\
        Old Exams & 94.74\% & 91.56\% & 82.49\%\\
        None of the above & 1.90\% & 1.06\% & 0.91\%\\
        \hline
    \end{tabular}
\end{table}

\begin{table}[h!]
    \caption{\label{tab11} Survey Question Five: Did you make any significant or intentional changes to your study methods for physics after using the supplemental resources provided by the department?}
    \centering
    \begin{tabular}{l   r   r}
    \hline
    \hline
        Answer Choice & Exam 1 & Exam 2\\
        \hline
        Yes & 53.34\% & 44.95\%\\ 
        No & 44.04\% & 52.33\%\\
        I did not use the resources & 2.61\% & 2.72\%\\
        \hline
    \end{tabular}
\end{table}

When asked to what degree students felt the supplemental materials impacted their performance on midterm exams, over 88\% of students reported that these materials had a positive impact on their exam performance and less than 5\% of students responded that they had a negative impact, Table \ref{tab12}. The last question posed to students was about whether or not they would recommend using these materials to their classmates and peers, Table \ref{tab13}, and over 98\% of students indicated that they would recommend utilizing the supplemental materials to other students.

\begin{table}[h!]
    \caption{\label{tab12} Survey Question Six: Did the supplemental resources provided by the department have a noticeable impact on your exam performance?}
    \centering
    \begin{tabular}{l   r   r   r}
    \hline
    \hline
        Answer Choice & Exam 1 & Exam 2 & Exam 3\\
        \hline
        Strong Positive Impact & 61.09\% & 55.00\% & 58.69\%\\        
        Slight Positive Impact & 30.57\% & 33.55\% & 30.06\%\\
        No Impact/Did Not Use & 6.40\% & 6.63\% & 6.96\%\\
        Slight Negative Impact & 1.36\% & 3.08\% & 3.32\%\\
        Strong Negative Impact & 0.58\% & 1.74\% & 0.98\%\\
        \hline
    \end{tabular}
\end{table}

\begin{table}[h!]
    \caption{\label{tab13} Survey Question Seven: Would you recommend the supplemental resources provided by the department to other students?}
    \centering
    \begin{tabular}{l   r   r   r}
    \hline
    \hline
        Answer Choice & Exam 1 & Exam 2 & Exam 3\\
        \hline
        Yes, strongly recommend & 88.08\% & 82.06\% & 86.15\%\\        
        Yes, somewhat recommend & 10.66\% & 16.30\% & 12.53\%\\
        No & 1.26\% & 1.64\% & 1.31\%\\
        \hline
    \end{tabular}
\end{table}

\section{Discussion}
Our regression models indicated that students' prior preparation in math and physics was highly predictive of average midterm exam performance. This is similar to prior studies where math preparation was the highest predictor of student learning \cite{long1986, hart1993, kost2009characterizing, tai2010, salehi2019, Stewart2021, meltzer2025pre} and that background physics knowledge was a less significant predictor of student performance \cite{meltzer2025pre}. In particular, we observed that students' recent mathematics preparation, based on final letter grade in Calculus I, was the single strongest significant predictor of their performance in a calculus-based introductory E\&M class at Texas A\&M. Prior physics knowledge and abilities, indicated by conceptual inventory pre-test scores were also a positive predictor of student performance, though carried less weight than math background, which is consistent with prior literature \cite{kost2009characterizing, salehi2019}. 

Beyond prior preparation, one of the most notable findings is the strong positive correlation between student performance and the use of supplemental old exams. This resource consistently emerged as the second strongest significant predictor of success, suggesting that students who engage with these materials can greatly benefit, regardless of their initial mathematics or physics background. This is an important result as more than 83\% of students indicated they used this resource according to anonymous surveys. As old exams offered only answers, not solutions, it is possible that students used this resource to test themselves prior to exams. Student use of repeated testing as a preparation strategy has been shown to positively impact performance, particularly if the exam or assessment does not immediately follow intensive studying (e.g. cramming) \cite{wheeler2003different, roediger2006test, karpicke2007repeated, mestre2020science}. Additionally, attempting prior versions of exams under test-like conditions can be effective in helping students to assess their level of preparation by providing an unbiased, objective measure of their abilities and identify areas for further study before taking exams \cite{mestre2020science}.

In contrast to prior studies on the impact of video resources, we do not observe positive correlations between students’ engagement with the materials and their course performance \cite{chen2010multimedia, sadaghiani2011, moore2018efficacyMLM}. Some methodological differences may account for the varied results. In the cited studies, students were assigned to engage with multimedia materials prior to class, and in at least one study, classroom instruction time was reduced commensurately with the length of the weekly videos. As a result, the videos were a more integral and required component of student learning, rather than optional supplemental materials. Our results also differ from those of Perry et al. \cite{perry2019newresource}, who observed that the supplemental concept videos included in their study had a positive impact, as measured by correlation coefficients, on student exam performance, particularly for the most conceptually challenging material (e.g., Faraday’s law) during the semester. However, none of these studies tracked student engagement with prior exams, an omission worth noting, as our analysis found this to be the second strongest predictor of student performance across all regression models. Linde et al. \cite{Linde2024} highlights survey findings on educators’ beliefs that supplemental resources foster the development of problem-solving skills and self-regulated learning. Through interacting with these materials, students can build the skills necessary to succeed not only within the course but throughout their academic and professional careers \cite{Singh2010olm}.

It's important to note again that the effect of using these supplemental materials appears to be even more pronounced for first-generation students compared to their continuing-generation peers. As seen in the first and third models, first-generation status on its own has a slightly negative correlation with students' exam performance. Interestingly, the beta values for the interaction term, representing first-generation students' use of old exams, are nearly equal in magnitude to the beta values for first-generation student status alone. This suggests that by utilizing this resource, students can minimize the observed disparity in exam performance. It is also possible that utilizing these supplemental resources may aid these students in developing crucial self-study skills. Prior research has shown that while accounting for general high school preparation can explain some differences in first-generation students’ course performance, the observed deficits cannot be attributed to mathematics and physics preparation alone \cite{Stewart2021}. This may be due to the fact that first-generation students often report greater challenges in managing their learning and possessing adequate study skills \cite{stebleton2012, falcon2015}. The implementation and further utilization of these supplemental materials could allow for all students, but especially first-generation students, to develop these crucial skills that are necessary for not just their academic future, but for their journey towards becoming life-long learners. 

An unexpected result was the consistent negative beta coefficient associated with student usage of supplemental chapter outline videos, which were aimed to guide students through reading of the textbook. One possible interpretation is that students relying more heavily on these videos are already struggling academically and seeking additional support in any form available. Alternatively, it is possible that some students are attempting to use these videos as their primary learning resource rather than as a supplement to reading as intended and in these cases, resource effectiveness can be reduced significantly \cite{deVore2017}. In such cases, their reliance on these materials may indicate a substitution effect, where they engage less with other learning strategies, such as attending lectures, participating in problem-solving exercises, or seeking peer collaboration. Prior research supports this interpretation, indicating that exclusively relying on online videos is less effective than attending regular lectures alone or using videos to complement classroom learning \cite{meehan2018}. Live instruction provides opportunities for real-time interaction, clarification, and deeper engagement with course content, elements that may be missing when students depend solely on passive video consumption. 

Another unexpected result was that two of our four supplemental resources, concept and problem-solving videos, showed no statistical significance within our models. However, student surveys revealed that over 50\% of students used the recitation-style problem-solving videos, and approximately 30\% utilized the supplemental concept videos. These resources also received high levels of approval, with many students indicating they would recommend them to their peers—suggesting a perceived positive impact on the learning process. 

Student perceptions gathered via surveys suggest that a large portion of students felt unprepared for their university-level physics courses after their completion of required high school curriculum. Over 60\% of students reported learning from online resources in general, however we note that a majority of students ($\geq$80\%) reported that our open access resources had a noticeable contribution to their learning outside of class. These results align with a central finding by Hill et al., which highlights that students who engaged with online learning materials alongside lecture tended to achieve greater conceptual gains compared to those who relied solely on traditional course instruction \cite{hill2015}. A similar proportion of students also claimed that usage of the supplemental resources had a tangible impact on their exam performance. A very encouraging result from our student surveys can be seen in student recommendations; an overwhelming amount of students, ($\geq$98\%), recommend the utilization of the department's open-access materials to their classmates and peers. This aligns with findings from anonymous surveys conducted by Perry et al. \cite{perry2019newresource} in spring 2017, in which students reported overwhelmingly favorable views regarding the impact of the supplemental concept videos.

Given the many complex and often unknown variables that influence student success across a course, these results highlight the importance of considering how students engage with different learning resources. Ultimately, our models offer important insight into the predictors of student success in introductory electricity and magnetism, helping to clarify the role of supplemental materials in shaping learning outcomes. 

\section{Limitations}
Although the regression models in this study indicate positive effects of student use of open-access supplemental resources, several limitations should be considered when interpreting the results. Usage data were based on the number of clicks on individual hyperlinks or visits to specific resource pages. However, the dataset did not capture time spent on each resource or actual engagement levels. Given the constraints of the website infrastructure, page visits were determined to be the most feasible proxy for resource utilization. Additionally, as previously noted, some students were excluded from the final regression models due to missing data—most commonly because calculus grades were only collected for students who completed the course in residence, or because some students did not complete the BEMA pre-test during the first recitation period.

\section{Conclusion}

We analyzed factors influencing student learning outcomes within an introductory calculus-based electricity and magnetism course, including demographics, prior preparation in mathematics and physics, and usage of open-access supplemental materials. We aggregated data from over 2000 students over three semesters to construct multiple linear regression models. Three models were run with approximately 1400 students for whom complete data were available. Consistent with prior studies, our results showed mathematical background, measured by students’ overall letter grade in their first-semester calculus course, was the strongest predictor of success, \(\beta\sim 0.314-0.327\). Importantly, across all three models, the second strongest predictor was student usage of supplemental old exams, \(\beta\sim 0.221-0.262\), which were used by a majority of students ($>$83\%). These relative beta values imply that student use of prior exams can significantly reduce the performance gap predicted by their level of math preparation. This could be due to them enhancing their self-regulated learning skills and continuously developing their problem-solving abilities. We observed that first-generation status was negatively correlated with student performance on exams. However, when combined with their usage of supplemental old exams, first-generation students showed a positive correlation with their performance, demonstrating that student use of this resource helps to bridge observed gaps. 

Data from our anonymous surveys indicated that two-thirds of students felt unprepared for this course after their high school physics experience. About half of students watched problem-solving videos and approximately one-third of students watched supplemental review videos. Responses further revealed that over 80\% of students who used the supplemental materials believed they made a noticeable contribution to their learning, and more than 88\% felt they positively impacted their performance. Remarkably, approximately 98\% of surveyed students recommended these materials to their classmates and peers.

These findings underscore the importance of accessible supplemental resources in supporting student populations with diverse backgrounds and levels of preparation---particularly those who face added challenges such as independently managing their learning. Resources like those described in this paper can help students develop the problem-solving and critical thinking skills essential for long-term success, and contribute to retention in STEM programs. As universities refine instructional strategies, integrating structured self-study tools, such as archived exams and instructional supplemental videos, may play an important role in promoting equitable outcomes in physics education. Future research should explore how variations in resource design and usage patterns affect long-term learning, ensuring that supplemental materials continue to evolve as effective and inclusive tools. Moving forward, we plan to conduct a similar analysis for our introductory calculus-based mechanics course.

\section{Acknowledgments}

We are grateful to Dr. A. Lewis Ford and Mr. Ryan Carmichael for their assistance with collecting and synthesizing the data for this study. This work was partially supported by Texas A\&M University and the Texas Higher Education Coordinating Board Digital Design for Student Success (D2S2) program.

\bibliography{InternalReportReferences.bib}

\end{document}